# On the role of secondary electrons in the color change of high-dose X-ray irradiated topaz

G. S. Elettivo[1,*], M. Ferraro[2,3,4,*,†], R. Filosa[2,3], A. Nicolino[2,3], B. Marmiroli[5], A. Turchet[6], and R. G. Agostino[2,3,4]

[1]*Department of Earth and Geoenvironmental Sciences, University of Bari Aldo Moro, Via Orabona 4, 70125 Bari, Italy*
[2]*STAR Research Infrastructure, University of Calabria, Via Tito Flavio, 87036 Rende, Italy*
[3]*Physics Department, University of Calabria, Via Pietro Bucci, 87036 Rende, Italy*
[4]*CNR Nanotec, Via Pietro Bucci, 87036 Rende, Italy*
[5]*Institute of Inorganic Chemistry, Graz University of Technology, Stremayrgasse 9/IV, 8010 Graz, Austria*
[6]*Elettra-Sincrotrone Trieste, SS 14 Km 163,5, AREA Science Park, 34149 Basovizza, (Trieste), Italy*

(Dated: January 16[th], 2024)

**\*** Correspondence: e-mail@e-mail.com

**Abstract:** Owing to its high brightness, synchrotron light allows for investigating with extreme precision the physical properties of matter. The irradiation with high-dose X-ray beams may also lead to modification of the latter, thus allowing for material processing. Here we investigate the color change of topaz irradiated with synchrotron light, shedding light on the role played by secondary electrons in the formation of color centers. As a matter of fact, treatments of natural topaz to induce its color change are largely used in the jewelry industry. Nevertheless, the physical mechanisms behind the topaz's color change have not yet been fully understood. To date, it has been shown that the combined action of high-energy beam irradiation (either electrons, neutrons, or γ-rays) and thermal annealing permits to provide colorless natural topaz with an artificial blue color, which is largely appealed in the gem market. Here we demonstrate that it is possible to irreversibly provide natural topaz with a blue color even by exploiting lower energy beams, such as X-rays, provided that enough dose is absorbed, thus paving the way for developing novel protocols for making artificially blue topazes.

# 1. Introduction

The advent of synchrotrons has been revolutionary across multiple scientific fields. Indeed, synchrotrons have transformed research by providing extremely bright and tunable X-ray beams, allowing physicists (and scientists in general) to probe the structure of matter with unprecedented precision. Synchrotron light has been primarily used for studying the structure and properties of materials at the atomic and molecular levels.

One of the advantages provided by synchrotron light is the fact that high-dose X-ray exposure can induce simultaneous synthesis and processing of materials. In the recent past, more and more attention has been drawn to the effect of irradiation on different materials; a remarkable example is the so-called X-ray lithography, which permits to create extremely fine patterns on substrates, e.g., for the production of high-precision, miniaturized components, driving advancements in semiconductor technology and nanofabrication [1].

In addition to fundamental research, X-ray beams, either soft or hard, are used to tailor the properties of materials for application-oriented purposes. In the realm of radiation-assisted synthesis and processing of materials, particularly interesting is the color change of minerals. A notable example is the irradiation of gems that makes them assume beautiful and appreciated colors, e.g., in the jewelry market. Remarkable examples of irradiated gems that are usually available in the market are quartz, corundum, tourmaline, and topaz. Generally speaking, minerals that are naturally colorless and whose crystalline structure contains hydroxyl (OH) groups may develop color centers after irradiation via particle or light beams as a consequence of $O^-$ hole center formation [2]. Interestingly, in most cases, the physical mechanisms behind the change of colors in minerals are not yet fully understood [3].

Here we consider the exemplary case of topaz, which crystallizes in the orthorhombic Pbmm space group [4] and has a very simple chemical structure; its chemical formula is $Al_2SiO_4(F,OH)_2$ as first independently determined by Pauling, Alston, and West in 1928 [5]. Owing to its stiffness, high refractive index, and low rate of inclusions, topaz is easy to polish, has an intrinsic brilliance, and is transparent. All these properties make it of particular interest for gemological purposes. Topaz is found in different colors in nature, from yellow to brown, azure, orange-pink, and red [6,7]. The latter is known as imperial topaz. Still, the large majority of natural topazes are colorless. It is for this reason that techniques for inducing a color change like irradiation have been developed. As a matter of fact, azure color-induced topazes are the most appealing in the gems market.

For all these reasons, among the minerals that change their properties as a consequence of irradiation, topaz is particularly suitable for investigating the origin of its change of color. The first paper that reported on irradiation treatment for varying the color of natural topaz dates back to 1947 when Pough showed that colorless topaz turns brown when irradiated by X-rays [8]. Since then, several papers investigated this effect [9-11]. To date, all X-ray-induced color changes in topaz have been shown to be reversible. Specifically, brown topazes recover their natural colorless look as a consequence of thermal treatments, e.g., heating up to about 300°C or exposing to sunlight [9].

At variance with X-rays, similar thermal treatments do not restore the topaz's original color when exposed to high-energy beams, e.g., fast electron and neutron or $\gamma$-ray beams. On the contrary, topazes acquire a bluish color, from dark blue (in that case topazes are labeled as "London") to azure, depending on the irradiation and thermal treatment parameters [12]. It is for this reason that, on the one hand, high-energy beam irradiation is largely used to produce commercial gems and, on the other hand, one may be led to believe that the color change is induced by effects that involve the physical effects on nuclear scale.

In this work, we report, we believe for the first time, that thermal treatment of X-ray irradiated topazes permits to provide them with a permanent (i.e., unchanged upon time in room ambient conditions) azure color. The novelty of our work is the use of a high-flux X-ray beam, i.e., synchrotron light. Whereas all studies of color changes in topaz induced by X-rays involved low-flux X-ray sources, e.g., in [11] topazes absorbed an X-ray dose of about 0.1 MGy. Here, we expose natural topaz samples to doses as high as hundreds of megagray. The achievement of blue topazes via X-ray irradiation demonstrates that the physical mechanisms behind the change of color are driven by electronic effects.

Specifically, in the context of radiation-induced effects in covalent solids, chemical bond modifications are induced by secondary electrons. In the "multiple-loss" model, primary photoelectrons and Auger electrons play a fundamental role in generating secondary electron emission. When X-ray photons are absorbed by atoms in the material, they eject primary photoelectrons, which, along with Auger electrons, initiate a cascade of inelastic scattering events. These electrons lose energy through interactions with other electrons (electron-electron scattering) and the crystal lattice (electron-phonon scattering). As a result, they excite additional electrons, creating a secondary electron avalanche [13]. The intensity of this avalanche, characterized by the number of secondary electrons generated per second per unit volume, is directly proportional to both the energy and the flux density of the incident x-rays, as higher x-ray energies and greater fluxes increase the rate and magnitude of secondary electron production. These electrons, produced by the direct interaction of ionizing radiation with the solid, own up to tens of eV and, as they traverse through the material, can transfer their energy to nearby atoms. Thus, the presence of secondary electrons in irradiated materials profoundly affects their physical and chemical properties through various mechanisms. The secondary electrons energy transfer to the material induces defect formation, atomic emissions, and structural changes. These effects are particularly significant in both wide-gap materials and semiconductors [14]. In particular, this energy transfer can cause bond breaking by providing sufficient energy to overcome the bond strength. A concurrent mechanism is the excitation of the atoms involved whose de-excitation pathways can lead to chemical bond destabilization. The overall effect leads to the formation of radicals or vacancies within the crystal lattice. The formers are highly reactive and can participate in chemical reactions, influencing the overall chemical structure of the material. On the other hand, defects can be created in the crystal lattice by displacing atoms from their regular positions. These defects, such as vacancies or interstitials, can significantly impact the material's properties and stability. Moreover, a low energy fraction (< 10 eV) of the secondary electrons is produced by the thermalization of the fast electrons. Those electrons act by promoting electron transfers or by interacting with neighboring atoms. This can lead to the formation of new chemical species and alterations in the material's chemical order.

Based on the analysis of transmission and absorption spectra at ultraviolet-visible (UV-VIS) and mid-infrared (IR) frequencies, respectively, we demonstrate that the interaction between secondary electrons and $OH^-$ groups plays a pivotal role in the change of color. The latter is accompanied by an irreversible reduction of the absorption threshold in the UV range that indicates the formation of a stable electronic state, clamped to the Fermi energy, which cannot be annealed by thermal effects. Our results, on the one hand, shed new light on the physical mechanisms behind the color change of natural gems and, on the other hand, will be of interest for applications in the field of jewelry.

## 2. MATERIALS AND METHODS

We used two natural colorless topaz samples originating from Tanzania and Mozambique, respectively. The first type comes from the Tunduru region, an area that is famous for its secondary deposits [15] which originate from the Kalahari formation and are rich in several minerals used as gems like sapphire [16]. Whereas Mozambican samples were extracted from a pegmatitic-origin primary deposit located in the Upper Ligonha [17]. In both samples, no inclusions were observed under the optical microscope. The two samples were prepared for irradiation and subsequent heating treatment as follows: from each sample, we prepared a plate having a thickness of about 1.1 mm with flat parallel faces and a shiny surface. The faces of the plates were parallel to the flaking plane.

All experiments were carried out at the Deep X-ray lithography synchrotron radiation (DXRL) beamline at the Elettra Synchrotron in Trieste, Italy [18]. In that setting, both topaz samples underwent irradiation using an intense X-ray beam characterized by a photon spectrum ranging from 1 to 30 keV (see the spectrum in Fig. 1).

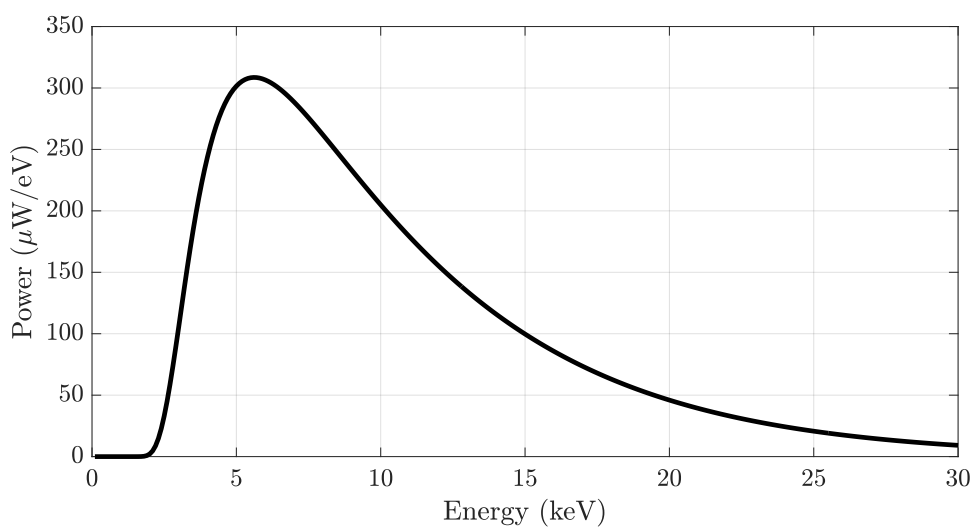

**FIG. 1.** Spectrum of the synchrotron beam used for irradiation.

To maximize the dose absorbed by the topaz no filters were applied. The fluxes during irradiation fell within the 1.7-3.2 W/cm² range, and the process was regulated with respect to time, ensuring a controlled deposited dose on the topaz samples within the range of tens of MGy to a few GGy.

In the irradiation campaign, the full synchrotron beam was sent through the topazes, from top to bottom. This permitted us to maximize the average dose absorbed by the samples (see Table 1). With the aim of making the topaz irradiation as uniform as possible, with the second exposure we flipped the sample; in this way, opposite sides of the plates directly faced the X-ray beam during the two irradiation steps. Due to the X-ray absorption of the topaz, in fact, the dose at the top side of the sample is higher than the one at the bottom since the beam intensity is quenched along the X-ray beam penetration according to the Lambert-Beer law. In the following, we will identify each step of the experimental process with the total average dose absorbed by the samples, i.e., the sum of the average dose absorbed during each round of irradiation. After irradiation, we perform a thermal treatment in air by means of a controlled temperature oven as reported in Table 1. We anticipate that the formation of the azure color was not observed after the annealing of fragments of the samples that were torn out after the first two expositions. We also emphasize that the choice of 260°C as the annealing temperature was made following the recipe of Ref. [19]. As shown in early studies, higher temperatures are likely to erase all the defects formed as a result of irradiation [9,19,20]. In contrast, lower temperatures may lead to a long annealing time to obtain the azure color.

**Table 1**. List of processes. Values of dose in the third column are the average between the two exposed sides (top and bottom).

| Process | Parameter | Value |
|---|---|---|
| First X-ray exposition | Dose (MGy) | 175 |
| Second X-ray exposition | Dose (MGy) | 350 |
| Third X-ray exposition | Dose (MGy) | 965 |
| Thermal treatment at 260°C | Time (min) | 30 |

Our samples were characterized using both UV-VIS (Cary 60, Agilent Technologies) and mid-IR (Alpha- T, Bruker Optics) spectroscopies. We used the following procedure: i) UV-VIS and mid-IR spectroscopy of the samples before irradiation; ii) Exposure of the sample to X-rays; iii) UV-VIS and mid-IR spectroscopy of the irradiated samples [the steps ii) and iii) were repeated until the sample became sufficiently opaque]; iv) Thermal treatment of the sample; v) UV-VIS and mid-IR spectroscopy of the heated samples [the steps iv) and v) were repeated until the sample became sufficiently transparent, possibly acquiring a bluish color].

## 3. Results

Before diving into the results of irradiated topaz it is useful to compare our natural samples with a London blue topaz that is commercially available. The blue color of the latter is likely to be achieved via neutron irradiation. Although no information neither in this regard nor on the origin of the gem was provided by the seller, neutron irradiation is the most common technique to make artificial London blue topazes. The merely qualitative comparisons in terms of the UV-VIS and Mid-IR absorption spectra of one of our natural colorless samples (that from Tanzania) and the commercial London blue topaz are shown in Fig. 2a,b, respectively. Some features are clearly detectable. First, the commercial blue topaz has a maximum of absorption around 600 nm (see Fig. 2a). As we shall see in detail in Section IV, the presence of an absorption band in the red spectral range is associated with a paramagnetic $O^-$ hole center at an OH lattice site (see [21] and reference therein).

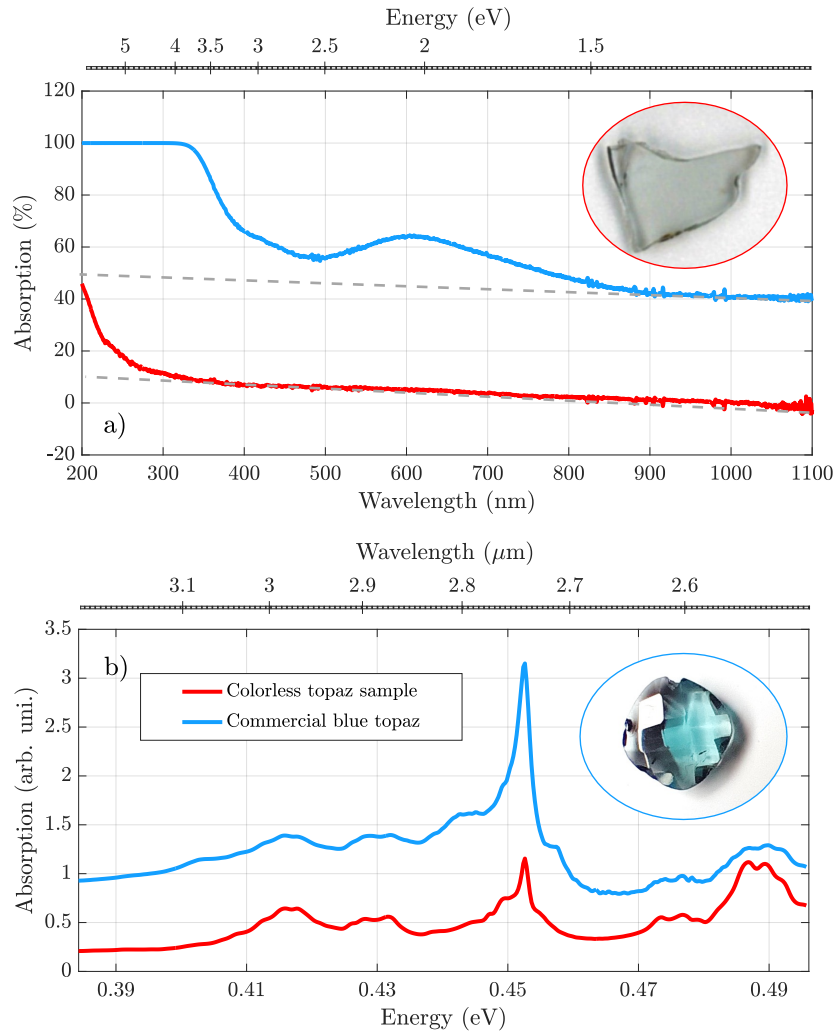

**FIG. 2.** Comparison between a commercial London blue topaz and one of our natural topazes: UV-VIS (a) and Mid-IR (b) absorption spectra, respectively. The insets in (a) are pictures of the two topaz samples. The gray dashed lines in (a) aim at emphasizing the slope of the background at large wavelengths, where topazes are transparent.

Moreover, the two samples have different bandgaps, which are identifiable at first approximation with the wavelength at which the absorption curve steeply approaches 100%. As can be seen in Fig. 2a, the London blue topaz's bandgap corresponds to a wavelength of around 350 nm, whereas that of the natural colorless topaz is well below 250 nm. Similar to the absorption of red light, the bandgap energy has been associated with OH-related defects [22].

Finally, in the mid-IR range, the London blue topaz has a marked absorption peak at 0.453 eV as shown in Fig. 2b. This is associated with the stretching of the $OH^-$ groups [23]. The same peak is also present in the absorption spectrum of the colorless topaz (cfr. red curve in Fig. 2b). However, the peak intensity is much lower than in the case of the London blue topaz: one may easily see that the peak at 0.453 eV has (does not have) the same intensity of the peaks between 0.484 and 0.496 eV in the case of the natural (commercial) topaz. In addition, the blue topaz shows two shoulders around the peak at 0.453 eV, characteristic of OH groups as well [23], which are originally absent in our colorless natural samples. These spectral features indicate that the $OH^-$ groups (and their associated defects such as hole centers) play a role in the transformation of a colorless to azure topaz, as shall be confirmed by the results on the irradiated samples reported below.

### 3.1 *UV-VIS spectra*

The effect of irradiation on the UV-VIS spectra of our samples is shown in Fig. 3. In order to better appreciate little variations of absorption, we used the log scale in the vertical axis. Moreover, in the horizontal axis, we indicate both the wavelength and the energy. Fig. 3a,b refer to Mozambican and Tanzanian samples, respectively. The modifications of the UV-VIS spectra are similar in all cases.

At first, X-ray irradiation produces a pronounced redshift of the absorption band at short wavelengths: the red curves in both panels of Fig. 3, which correspond to not-yet-irradiated samples, are spectrally separated from all the other curves. In particular, the absorption threshold after irradiation is $E_g \sim 3.2$ eV.

Another main spectral feature associated with X-ray irradiation is the emergence of an absorption peak around 2 eV, i.e., in the red spectral range. It can be noticed that the intensity of such a peak increases as the absorbed dose grows larger (cfr. orange, yellow, and green curves in Fig. 3).

Remarkably, both the absorption threshold redshift and the formation of a pronounced absorption in the red spectral range are irreversible (cfr. red and blue curves in Fig. 3). Even after thermal treatment, in fact, the absorption band at short wavelengths does not recover its initial spectral position. Analogously, the absorption peak in the red spectral range is not quenched by the thermal annealing. It is interesting to notice that two "seeds" of absorption near 2 and 3.5 eV were already present before irradiation (see red curve in Fig. 3a). This somehow indicates that these features are related to stable electronic states, as will be discussed in Sec. IV.

On the other hand, thermal annealing strongly reduces the absorption in the blue-to-green spectral range, which is associated with a plethora of metastable states. This provides the sample with renewed transparency so that the topaz turns from brown to azure, provided that enough X-ray dose is absorbed. Both the absorption threshold redshift and the color variation will be discussed in detail in the following.

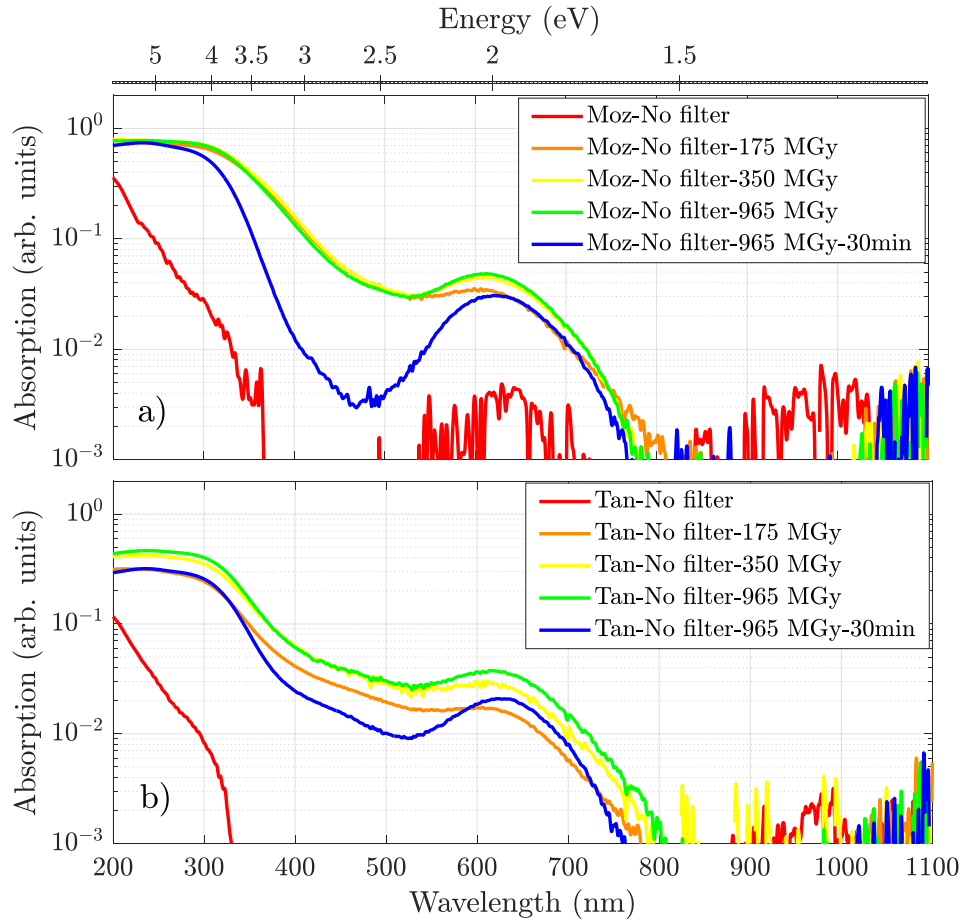

**FIG. 3.** UV-VIS spectra of irradiated topazes: (a,b) Mozambican (a) and Tanzanian (b) samples that have undergone X-ray irradiation. The values in the legend refer to the total average absorbed dose.

### 3.2 *Absorption threshold redshift*

Let us first focus on the details of the absorption threshold redshift. An accurate evaluation of the bandgap value was obtained as follows: we considered the UV-VIS absorption spectra after removing the linear background, indicated as a dashed line in Fig. 2a, i.e., we forced the absorption spectra to approach zero at large wavelengths, in the nominal transparency window of topaz. For the sake of readability, we reported in Fig. 4a the results of the Mozambican sample. Specifically, we compare the absorption spectrum of the sample before any irradiation (red curve) and after the second thermal treatment (blue curve). The material absorption threshold was found as the intersection of the horizontal axis, i.e., zero absorption condition, with the linear fit of the spectrum at its inflection point (dashed gray line).

The absorption threshold redshift of both the Tanzanian and Mozambican topazes as a consequence of the absorption of X-rays is shown in Fig. 4b. As can be seen, a drop of $E_g$ of about 2 eV occurs even at the lowest dose; a further decrease, albeit much smaller, is observed as the X-ray dose grows larger. The absorption threshold energy shift turned out to be practically insensitive to the thermal treatment. Indeed, we found $E_g \sim 3.4$ eV regardless of the heating time (not shown).

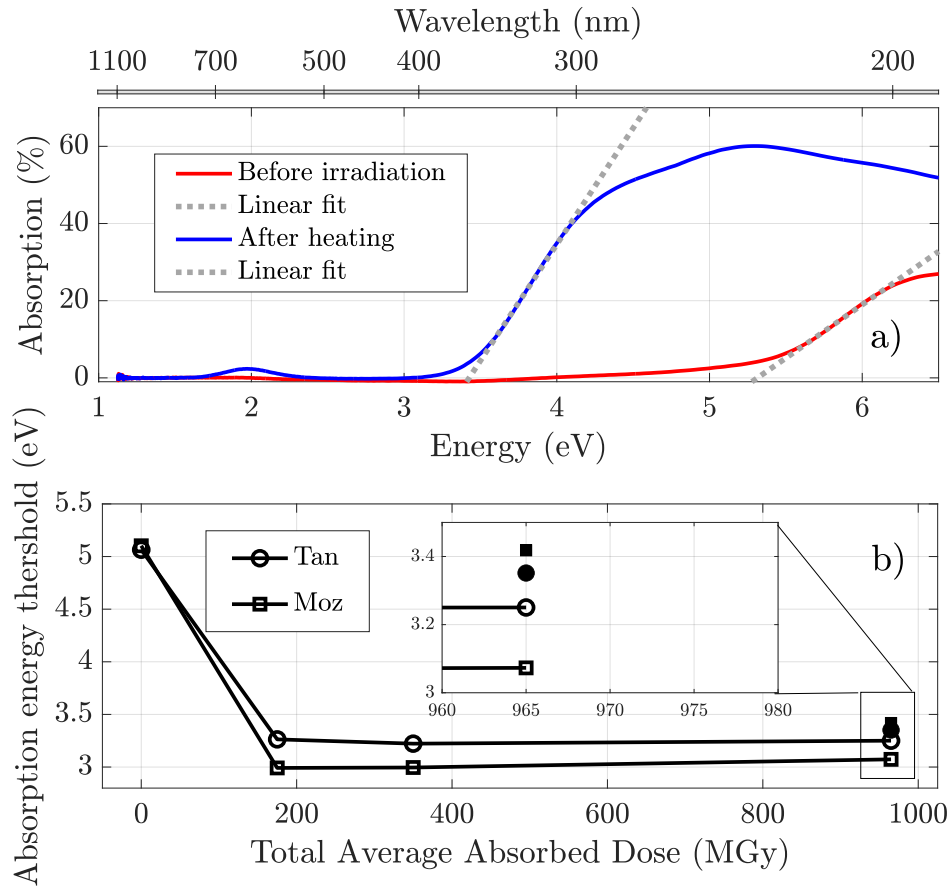

FIG. 4. Absorption threshold redshift. (a) UV-VIS absorption spectra of a Mozambican sample before X-ray irradiation and after thermal treatment as in Table 1. (b) Absorption threshold vs. absorbed dose. The inset shows the effect of thermal treatment, indicated by arrows.

### 3.3 *Color variation*

The most relevant result for application purposes is definitively the topaz color change. In order to stress this aspect, we report in Fig. 5a a comparison of the absorption curves in the visible spectral range of the Mozambican sample, before the first irradiation, i.e., untreated, after the last irradiation, and after the thermal annealing. Within this fashion, one can better appreciate that the color change from colorless to azure occurs because of two effects. the formation of an absorption peak at about 620 nm (2 eV), which makes the topaz absorb red light, and the quenching of the absorption band below 500 nm, which, instead, makes the topaz transparent to the blue light. To further emphasize these aspects we add an inset on the top of Fig. 5a representing the visible light spectrum.

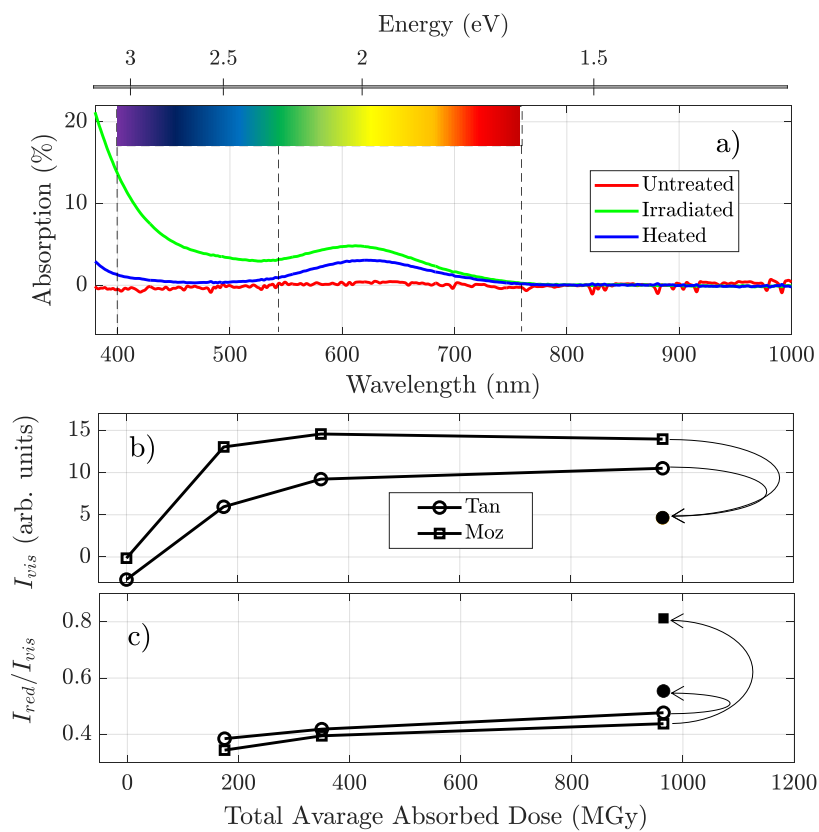

FIG. 5. Topaz color variation. (a) Visible absorption curves of the Mozambican sample. The inset on the top of the panel represents the visible spectrum. The gray dashed lines represent the intervals of integration that define $I_{vis}$ and $I_{red}$. (b,c) $I_{vis}$ (b) and ratio $I_{red}/I_{vis}$ (c) vs. absorbed dose. Full symbols indicate data of samples after thermal annealing.

A quantitative analysis of the topaz color variation is reported in Fig. 5b,c. We calculated the values of the integral of the absorption curves in Fig. 3 both over the whole visible spectral range (400-750 nm) and over the red part of the spectrum (550-750 nm). These integrals are dubbed as $I_{vis}$ and $I_{red}$, respectively. The integration intervals are indicated by vertical dashed lines in Fig. 5a.

In Fig. 5b,c, we plot the values of $I_{vis}$ and the $I_{red}/I_{vis}$ ratio as a function of the X-ray dose. As can be seen, although the visible absorption increases as the dose grows larger, the $I_{red}/I_{vis}$ ratio remains nearly the same during the irradiation processes in the case of the Tanzanian sample. This is the origin of the brown color assumed by the topaz induced by X-ray irradiation. On the contrary, the thermal treatment quenches the visible absorption, i.e., the samples become overall more transparent to visible light, and, at the same time, it enhances the red-to-visible absorption ratio, i.e., the samples turn bluish. Such a double mechanism that leads to the topaz change is even more evident at higher exposition doses. This can be seen in Fig. 5b,c; the effect of thermal treatment on the Mozambican sample made the $I_{red}/I_{vis}$ ratio almost double (see arrows in Fig. 5c), while X-ray irradiation practically produces a uniform increase of the absorption in the red and green-blue color ranges.

### 3.4 *Mid-IR spectra*

As shown in Fig. 2b, mid-IR absorption spectra provide information about the nature of the physical mechanisms that lead to the topaz color change. For the sake of simplicity, in Fig. 6a we reported only the spectra of the Tanzanian sample before any irradiation, after the last irradiation round, and after thermal annealing. The main features of the spectra were the same for all the samples considered in this work. In order to emphasize the modification induced in the absorption at mid-IR frequencies by the X-ray irradiation and the thermal treatment, in Fig. 6b, we reported the difference between the absorption curves in Fig. 6a.

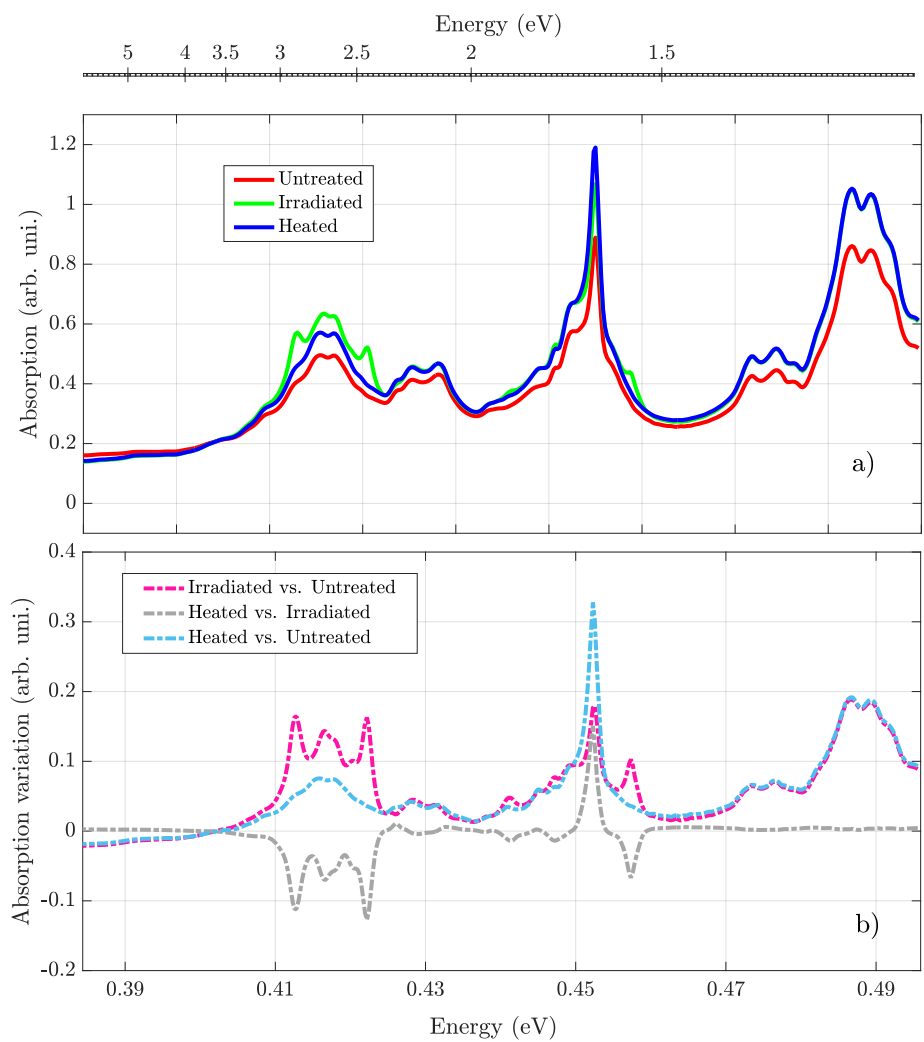

FIG. 6. Mid-IR absorption spectra. (a) Absolute absorption. (b) difference between the spectra in (a).

As can be seen, X-ray irradiation produces the enhancement of the absorption at specific spectral positions. Indeed, the magenta curve in Fig. 6b is characterized by several peaks: a cluster of peaks between 0.409 and 0.421 eV, which is due to the stretching of the anomalous OH groups correlated to the brown color as reported by Aines and Rossman in [24], a main resonance at 0.452 eV accompanied by side peaks, i.e., the fingerprint of the stretching of the OH groups, and two large lobes between 0.484 and 0.496 eV. The small peak at 0.455 eV belongs to OH-bearing phases as reported by Wunder et al. [25]. The sharp peak at 0.452 eV, instead, is related to OH groups and, specifically, to those having fluorine in the opposite site; whereas, the shoulder right below 0.452 eV is caused by two opposite OH groups [23]. In this regard, it is worth mentioning that the coexistence of these two naturally occurring crystallographic defects associated with an $O^-$ ion on an (OH) site was further confirmed by electron paramagnetic resonance [26,27] as well as magic angle spinning nuclear magnetic resonance and Raman spectroscopy measurements [28].

Interestingly the thermal annealing produces a nontrivial variation of the absorption curve (gray curve in Fig. 6b). The spectral variations induced by X-ray irradiation between 0.409 and 0.421 eV are almost "quenched" by the annealing, in analogy to the visible absorption between 400 and 550 nm in Fig. 5a. The same occurs for the peak at 0.455 eV. Whereas the main resonance at 0.452 eV is further enhanced by the thermal treatment; in analogy to the $I_{red}/I_{vis}$ ratio in Fig. 5b,c. Remarkably, our observation is opposite to that of Ittipongse et al [22], who noticed a decrease in the intensity of the peak at 0.452 eV, ascribable to the decomposition of the OH bonds in favor of the growth of $O^-$ and $H_0$ vacancies, i.e., hole and electron traps, respectively. On the other hand, the spectral features around 0.449 eV, associated with the OH-OH coordination, survive thermal treatments, just like the absorption threshold redshift discussed in Sec. III.B. Similarly, the two lobes at large energies are insensitive to the annealing.

As such, the combined action of irradiation and thermal treatment results in the enhancement of the absorption associated with the resonance at 0.452 eV, its shoulder at lower energy, and the lobes between 0.484 and 0.496 eV (cfr. the cyan curve in Fig. 6b).

**4. Discussion**

We now discuss, in light of the results reported in the previous section, the physical mechanisms that lead to the topaz color change.

First, it is important to mention that some of the features related to the irradiation of topaz have already been reported in the literature. For instance, the redshift of the absorption band in the UV to 3.4 eV has been associated with the increase in the density of defects that are formed during irradiation [22]. The absorption band between 400 and 550 nm, instead, has been ascribed to $H^{2+}$ and $H^{3+}$ electronic traps [29]. Finally, the absorption band around 620 nm (2.0 eV) has been largely studied in the framework of radiation-induced color centers: the presence of $O^-$ centers (hole centers) that interact with two adjacent Al ions was indicated as its possible origin [30]. In this regard, it is worth mentioning that Da Silva [30] and Priest [31] showed that the color change, and thence the absorption peak at 620 nm, cannot be due to the $O^{2-}$ peroxyl radicals because these have higher thermal stability than the $O^-$ centers. Whereas, relying on thermoluminescence measurements, Isotani et al. found that the optical absorption at 620 nm can be due to two different structural configurations of $O^-$ states, which were associated with polarons adjacent to electron traps [21].

For the sake of readability, we report in Fig. 7a the absorption spectrum in the linear scale of the Tanzanian sample. In Fig. 7b, we show the detail of the curves in Fig. 7a around the red absorption peak.

Let us first consider the reduction of the absorption energy threshold in the UV-VIS spectral range; in addition to results in Sec III.B, we found that such a reduction occurs even at doses as low as 1 MGy (not shown). However, topazes exposed to high doses turned azure after annealing, while it is well-known that color changes due to low-dose X-ray exposition are reversible via thermal treatments [8-12]. This indicates, on the one hand, that the bandgap reduction, albeit irreversible, is not sufficient to ensure the achievement of the azure color and, on the other hand, that an electronic state, stable to thermal heating, is formed as a consequence of X-ray irradiation (regardless of the dose).

Via thermal annealing, instead, an abundance of metastable states relaxes. This is indicated by the quenching of the absorption curve in the blue spectral range (cfr. Fig. 7c, where we plot the difference between the absorption curves in Fig. 7a after heating and irradiation, respectively).

The stable state is formed within the material original bandgap by material Frenkel defects, such as $O^-$ centers induced by secondary electrons. Moreover, the stable state is related to the OH groups, as demonstrated by the mid-IR spectra illustrated in Section III.D. Specifically, our results indicate the occurrence of structural modifications, in favor of OH-OH coordination, as a consequence of X-ray irradiation. This is consistent with studies of the OH mid-IR spectrum variation with both temperature [23,32] and pressure [33]. Thus, our results indicate that secondary electrons produce an overall increase of both F and OH neighboring in the F/OH site, which has a non-linear effect on the expansion or contraction of the crystal structure [23]. This is proven by the simultaneous increase of both the OH stretching bands in the IR spectra of topaz, i.e., the peak at 0.452 eV and its shoulder at lower energy in Fig. 6 (cyan curve). The absorption at these frequencies, in fact, is related to the local surroundings of the F/OH site in the crystal structure. As such, one may infer that the lattice distortion induced by X-ray irradiation serves to stabilize a cluster comprising several equivalent oxygen ions, effectively trapping the $O^-$ hole located adjacent to an OH Frenkel defect that plays the role of an acceptor. This is consistent with models of the effect of doping and ion implementation in topaz, which showed that OH Frenkel defects have the lowest formation energy threshold [34].

Here we found that the absorption curve after thermal annealing, subtracted by that of the untreated sample, has a symmetric bell shape, which is well fit by a Gaussian function (see Fig. 7d). This is related to the joint density of state between the valence band and the electron state associated to the defect. Now, the analysis of the band structure of the topaz either before or after irradiation is a complex task that goes beyond the scope of this work. However, in Fig. 7e, we present a schematic energy diagram of the azure topaz, providing a speculative overview of the electronic aspects corresponding to the observed optical changes. The electron state is placed at an energy $E_2$ above the valence band. Here $E_2$ = 2.0 eV, i.e., the energy at which the absorption curve of the topaz has a maximum in the red region. Whereas the energetic gap between the stable state and the conduction band, namely $E_1$, is given by the spectral position of the redshifted absorption threshold, i.e., ~3.2 eV. The validity of the picture is confirmed by the fact that $E_1 + E_2 = E_g$. Finally, the presence of a peak at energy $E_2$ in the absorption spectrum indicates that the electronic state is only partially occupied, i.e., it is clamped to the Fermi energy $E_F$ as in Fig. 7e.

The X-ray dose somehow determines the occupation of the stable state. Therefore, if the dose is too low the material only experiences a reduction of its absorption threshold in the UV-VIS spectral range as the stable state is formed but its occupancy does not favor the absorption of photons with energy $E_2$.

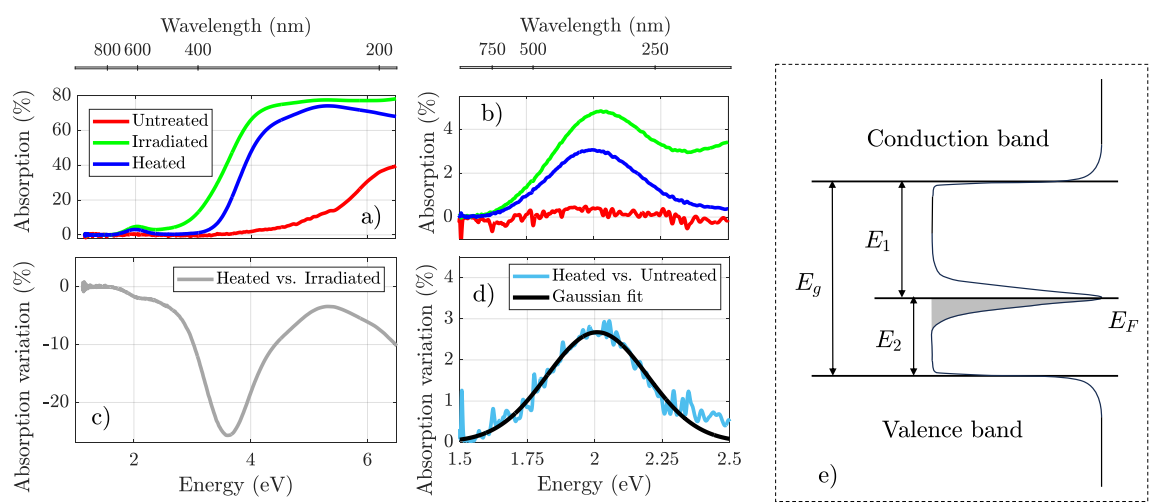

FIG. 7. (a) Absorption curves of the Tanzanian sample. (b) Zoom of (a). (c) Difference between the blue and the green curves in (a). (d) Difference between the blue and the red curves in (b). (e) Sketch of the energy levels of the topaz after thermal annealing. The solid curve represents the density of states; the gray zone indicates the electron occupancy of the state within the material bandgap, being $E_F$ the Fermi energy.

## 5. Conclusion

In conclusion, we carried out a study of the color change to azure of natural topaz as a consequence of X-ray exposition of hundreds of megagray and thermal annealing at 260 °C. The achievement of the azure color via X-ray irradiation demonstrates that pure electronic effects drive the physical mechanisms behind the change of color.

Based on UV-VIS and mid-IR spectroscopy we demonstrated that the color change is ascribable to material defects, which are related to OH⁻ groups. and originate from the production of secondary electrons. In particular, O⁻ centers are responsible, on the one hand, for non-negligible absorption of red light and, on the other hand, for an irreversible absorption energy threshold reduction in the UV spectral range. These effects are irreversible; whereas thermal annealing permits to quench the absorption in the green-blue spectral window, i.e., it induces the relaxation of a plethora of metastable states, thus providing topaz with a beautiful azure color, which, besides fundamental scopes, is particularly appealing for jewelry-related purposes. Finally, as. proof of the durability of the color change we would like to mention that the irradiated and annealed samples were stored for about three years in standard environmental conditions, i.e., at room temperature and atmospheric pressure, without showing any loss of their azure pigmentation.

**Acknowledgments**

We acknowledge the financial support of the Ministero dell'Istruzione, dell'Università e della Ricerca grant STAR-2 (PIR01-00008). The access to DXRL beamline experimental facilities at Elettra-Sincrotrone